# Temperature and strain characterization of regenerated gratings

Tao Wang,[1,2] Li-Yang Shao,[1,3] John Canning,[1,*] Kevin Cook,[1]

[1] *Interdisciplinary Photonics Laboratories (iPL), School of Chemistry, The University of Sydney, NSW, 2006 Australia*
[2] *Institute of Optoelectronic Technology, Beijing Jiaotong University, Beijing 100044, China*
[3] *Department of Electrical Engineering, The Hong Kong Polytechnic University, Hung Hom, Kowloon, Hong Kong SAR*
*\*Corresponding author: john.canning@sydney.edu.au*

Both temperature and strain characterization of seed and regenerated gratings with and without post-annealing is reported. The high temperature regeneration has significant impact to thermal characterization and mechanical strength of gratings whilst the post annealing has little effect. The observed difference is evidence of viscoelastic changes in glass structure.

There is an obvious upper limit to environmental temperature monitoring using conventional fiber Bragg grating (FBG) sensors. They have been developed for telecommunications and optimised to operate between −20 to + 80 °C for up to 25 years. Over shorter durations they are capable of working up to 200 °C, though they tend to anneal rapidly above 300 °C [1]. These type I gratings are commonly inscribed within germanosilicate optical fibers (e.g. Corning SMF-28) or boron co-doped germanosilicate fibers to enhance photosensitivity whilst matching numerical aperture with SMF-28. Regenerated gratings are a recent addition to the grating menagerie, optimised for increasingly higher temperatures, above 1100 °C [2-4]. High temperature stability opens up access to applications in harsh environments where conventional type I gratings cannot be used. For instance, they have been used to characterize the high temperature environment within an MCVD tube [5], for developing high temperature pressure sensors [6], and used to measure exhaust temperatures from diesel train turbines [7].

In this letter, the temperature and strain properties of regenerated gratings are characterized. Regeneration involves a greater physical contribution through structural relaxation at areas where stresses are altered in the presence of hydrogen, which simply by being present acts to reduce tensile stresses across the core and cladding and between processed regions – a recent review reveals new insight in this regard [8]. The local relaxation differs between areas of high and low laser exposure during fabrication of the seed grating. Hydride and hydroxyl formation helps to accentuate this difference, which leads to stronger grating writing as well as altering stresses [9], and therefore local pressures, periodically.

Bragg gratings were inscribed in B-codoped germanosilicate fiber ([$GeO_2$] ~ 33 mol%; [$B_2O_3$] ~ 12 mol %) by direct writing through a 10mm optical phase mask using an ArF laser ($\lambda$ = 193 nm; pulse fluence: $f_{pulse}$ = 95 mJ/cm$^2$; cumulative fluence $f_{cum}$ = 113 J/cm$^2$; $RR$ = 30 Hz; pulse duration $\tau_w$ = 15 ns). Before grating writing, the fiber was hydrogen ($H_2$) loaded ($T$ = 80 °C, P = 180 bar, $t$ = 4 days), avoiding type I$n$ formation [10]. For regeneration, two groups of seed gratings were fabricated - Group (1): regenerated gratings, and Group (2): regenerated gratings subjected to post-annealing at 1100 °C.

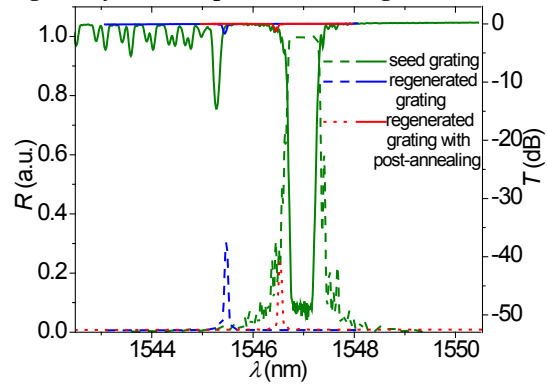

Fig. 1. Initial reflection (dash) and transmission (solid) spectra of a type I seed grating with $Tr$ = 50 dB, and final spectra of regenerated grating with / without post-annealing.

The typical initial spectra with both reflection, $R$, and transmission, $Tr$, of the seed gratings are shown in Fig. 1. A high-temperature heater was used for annealing the gratings and temperature monitored with a type $K$ thermocouple. The thermal processing recipe is shown in Fig. 2 along with the grating reflection strengths (normalized to the initial seed grating). The temperature rose to $T$ = 850 °C over $t$ = 60 min before dwelling for $t$ = 40 min. Over this period, the grating decays completely before regenerating and saturating at a peak reflection. At the beginning of the thermal process, a growth in the reflectivity is observed indicating an annealing out of a negative contribution, possibly arising from stress, early on has occurred, reminiscent of behavior observed with type I$n$ (type IIA) formation and annealing. For Group (2) gratings, additional annealing ($T$ = 1100 °C, $t$ = 20 min) was undertaken to stabilise the regeneration process for even higher temperature operation, shown in Fig. 2. The strength of the grating decays further before stabilizing. Fig.1 shows the final spectra of the regenerated gratings.

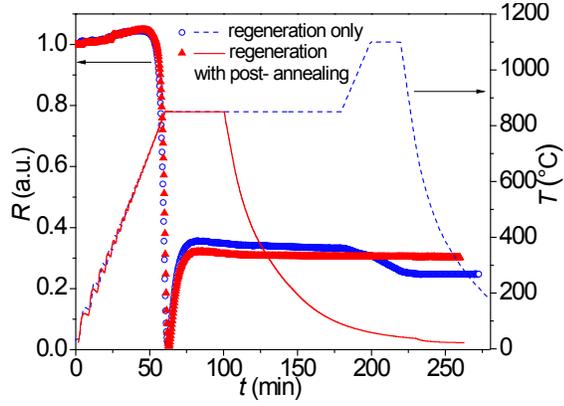

Fig. 2. The regeneration process with / without post-annealing: seed gratings are heated from $t = 0$. Temperature profile is also shown (red curve and blue dash curve).

If there is no strain and only the temperature changes, the Bragg wavelength, $\lambda_B$, changes as [11]:

$$\Delta\lambda_B = \lambda_B (\alpha + \xi) \Delta T \quad (1)$$

where $\alpha = (1/\Lambda)(d\Lambda/dT)$, $\xi = (1/n_{eff})(dn_{eff}/dT)$, and

$$\Delta n = \xi\, n_{eff}\, \Delta T \quad (2)$$

$\xi$ = thermo-optic coefficient, $\alpha$ = thermal expansion coefficient. For fused silica $\alpha = 5.5 \times 10^{-7}$ °C$^{-1}$ [12].

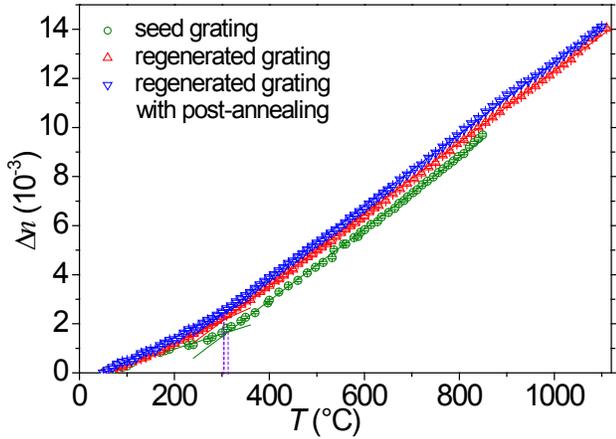

Fig. 3. $\Delta n$ versus $T$ for each grating type.

The temperature dependence of the regenerated gratings was characterized using a broadband source and an optical spectrum analyzer over the temperature range 50 to 1100 °C. The data of the seed gratings is from 50 to 850 °C because it decayed completely at 850 °C. By monitoring the shift in wavelength, $\Delta\lambda_B$, $\xi$, $d\lambda_B/dT$ and the average index change, $\Delta n (T)$, can be obtained through (1)-(2). Fig. 3 shows $\Delta n$ versus $T$ for the three types of gratings. $\xi$ and $d\lambda_B/dT$ are as shown in Table.1. Different gradients are observed below and above $T = 300$ °C defining a low and high temperature regimes. Above 300 °C, each grating has similar values for $d\lambda_B/dT$ and $\xi$. Within experimental error, these results indicate that subsequent high $T$ annealing does not alter the core properties further. In the low temperature regime, the seed grating has the lowest value for $d\lambda_B/dT$ and $\xi$, regenerated gratings are higher and post-annealed regenerated gratings have the highest value. This is the evidence that high $T$ processing and annealing during fabrication has significant impact on the subsequent temperature response of gratings at low $T$ but not at high $T$, consistent with significant glass relaxation.

To compare with existing material systems, for silica $\xi \sim 7 \times 10^{-6}$ °C$^{-1}$ has been reported [13] in agreement with bulk estimates $\sim 7.2 \times 10^{-6}$ °C$^{-1}$ [14]. Through simple linear superposition of constituent elements within the core, the addition of Ge should raise $\xi$ which is usually observed. For example, measured using long period gratings $\xi \sim 7.6 \times 10^{-6}$ °C$^{-1}$ [15], slightly higher than silica. However, at high temperatures, $\xi \sim (9.90 \pm 0.05) \times 10^{-5}$ °C$^{-1}$ noticeably higher than either pure fused silica or germanosilica.

On the other hand, the presence of B has the reverse effect to that of Ge, where a small reduction in $n$ by adding B also leads to a small reduction in $\xi$ at lower temperatures [16]. Generally, the magnitude of the change in $n$ adding B is much less than Ge. Therefore, to reduce the rises in $n$ obtained with Ge, higher concentrations of B are often used. By contrast, the $\alpha$ of borosilicate is significantly higher for similar concentrations. A greatly reduced $\xi \sim (5\text{-}6) \times 10^{-6}$ °C$^{-1}$ [17] was obtained in previous measurements for type I gratings in fibre similar to that used in this work. This value is between our seed grating and post-annealed regenerated grating at low temperature, and consistent with the regenerated grating. There is a remarkable change above a threshold-like temperature $T \sim 312$ °C for the seed (Fig. 3), which is similar to both types of regenerated gratings where $T \sim 304$ °C. The sudden onset of this change would preclude a simple change in modal overlap. Rather, we propose that this transition in effect marks the glass transition temperature, $T_g$, for the borate constituent, noting that $T_g$ for borate is highly dependent on borate concentration [18], varying between 250 °C to 570 °C, for example [19].

Table 1. $T$ response of $\lambda_B$ and $\xi$ for each grating.

| Grating type | Section of $T$ (°C) | $d\lambda_B/dT$ (pm °C$^{-1}$) | $\xi$ (10$^{-6}$ °C$^{-1}$) |
|---|---|---|---|
| Seed | 50-300 | 7.4 ± 1.6 | 4.2 ± 1.0 |
| | 300-850 | 16.2 ± 0.9 | 9.9 ± 0.6 |
| Regenerated | 50-300 | 10.1 ± 0.5 | 6.0 ± 0.3 |
| | 300-1100 | 16.1 ± 0.1 | 9.9 ± 0.1 |
| Regenerated & Post-annealed | 50-300 | 11.0 ± 0.9 | 6.5 ± 0.6 |
| | 300-1100 | 16.3 ± 0.5 | 9.9 ± 0.3 |

To characterize the response of the gratings with longitudinally applied strain, the fiber grating is fixed, using epoxy, between two displacements stages. The fiber is stretched with a calibrated micrometer and measurements of both $\Delta L$ and $L$ are taken. The initial tension is carefully fixed and then increased by turning the micro positioner. Each increment length change and the corresponding $\lambda_B$ were recorded. Fig. 4 summarizes the results. For all three types of gratings the observed shift in $\lambda_B$ and applied strain is linear: $d\lambda_{B\,seed}/d\sigma = (9.83 \pm 0.15) \times 10^{-4}$ nm/$\mu\varepsilon$. This is higher than values reported for other germanosilicate fiber where $d\lambda_B/d\varepsilon = 7 \times 10^{-4}$ nm/$\mu\varepsilon$ at lower temperatures [17], though it is consistent with other work in similar B doped (softened) germanosilicate fiber [20]. For two types of regenerated gratings, the response of $\lambda_B$ are commensurate: $d\lambda_{B\,reg}/d\varepsilon = (9.61 \pm 0.16) \times 10^{-4}$ nm/$\mu\varepsilon$ and $d\lambda_{B\,anneal}/d\varepsilon = (9.62 \pm 0.23) \times 10^{-4}$ nm/$\mu\varepsilon$. If the temperature is constant, $\lambda_B$ changes with strain as [11]:

$$\Delta\lambda_B / \lambda_B = (1-p_e)\,\varepsilon \qquad (3)$$

$p_e$ is an effective strain-optic constant:

$$p_e = (n^2/2)[p_{12} - \nu(p_{11}+p_{12})] \qquad (4)$$

where $p_{11}$ and $p_{12}$ are components of the strain-optic tensor and $\nu$ is Poisson's ratio. For the strain applied in this experiment (<1300 $\mu\varepsilon$), the change of $p_e$ is <0.1%. The difference between regenerated gratings and seed grating indicates changes in the strain optic components also consistent with relaxation of the glass.

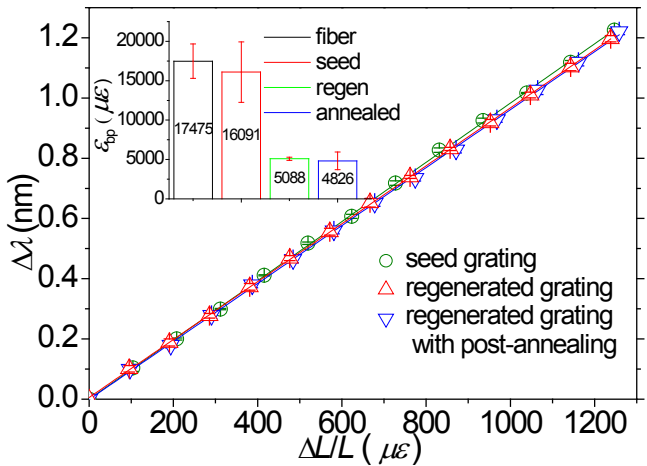

Fig. 4. $\Delta\lambda$ versus $\Delta L / L$ for three types of gratings. Inset: Breaking point for three types of gratings and for pristine fiber.

With sufficient load, the breaking points of the three grating types were measured. Pristine hydrogen loaded samples of the same fiber are also tested for reference. The inset of Fig. 4 shows the breaking point of seed gratings is $\varepsilon_{bp} = 16091$ $\mu\varepsilon$, close to the H$_2$ loaded pristine fiber (no UV exposure) $\varepsilon_{bp} = 17475$ $\mu\varepsilon$. The values for regenerated and post-annealed regenerated gratings are $\varepsilon_{bp} = \sim 5088$ $\mu\varepsilon$ and 4826 $\mu\varepsilon$, which is about 1/3 of seed gratings. This means the high temperature regeneration process reduces the mechanical strength of gratings; in part this may be due to H$_2$ reactions and fracturing as well as water ingress mad cracks probably at surface scratches caused by fiber stripping. The post annealing itself generates no additional degradation.

In summary, both the temperature and strain of seed, regenerated and post-annealed regenerated gratings have been characterized. All three types of gratings have a threshold-like temperature near $T \sim$ 310 °C which is attributed to the low glass transition temperature of borate. The high temperature processing and annealing have significant impact on the thermal characterization of gratings at low $T$ but not at high $T$. All three types of gratings have a linear response in $\Delta\lambda_B$ with applied strain. The regenerated gratings have slightly lower values than the seed gratings. The high $T$ regeneration degrades the mechanical strength of the seed gratings whilst the post annealing has little effect to that of the regenerated gratings. The results are consistent with glass relaxation as the causal effect of regeneration.

Australian Research Council (ARC) FT110100116 funding is acknowledged. T. Wang acknowledges a China Scholarship Council (CSC) Visiting Scholar Award. L. Shao acknowledges an Australia Award Endeavour Research Fellowship, HK. Poly. Univ. project G-YX5C and NSFC Grant No. 61007050.